\documentclass[10pt]{article}
\usepackage{latexsym}
\usepackage{amssymb}
\usepackage{amsmath}
\usepackage{amscd}
\usepackage{amsthm}
\usepackage[left=2cm,top=2.5cm,right=2.0cm,bottom=2.0cm]{geometry}
\usepackage[dvips]{graphicx}
\usepackage{hyperref}
\usepackage{textcomp}
\begin{document}
\begin{center}
\Large{\bf{A transitioning universe with anisotropic dark energy}}\\
\vspace{4mm}
\normalsize{Anil Kumar Yadav}\\
\vspace{4mm}
\normalsize{Department of Physics}\\
\vspace{2mm}
\normalsize {United College of Engineering $\&$ Research, Greater Noida - 201306, India}\\
\vspace{4mm}
\normalsize{E-mail: abanilyadav@yahoo.co.in}\\
\vspace{2mm}
\end{center}
\begin{abstract}
In this paper, we present a model of transitioning universe with minimal interaction between perfect fluid and 
anisotropic dark energy in Bianchi I space-time. The two sources are assumed to minimally interacted and therefore 
their energy momentum tensors are conserved separately. The explicit expression for average scale factor are considered 
in hybrid form that gives time varying deceleration parameter which describes both the early and late time physical 
features of universe. We also discuss the physical and geometrical properties of the model derived in this paper. 
The solution is interesting physically as it explain accelerating universe as well as singularity free universe.   
\end{abstract}
\section{Introduction}
The astronomical observation of SN Ia \cite{riess1998, perlmutter1999}, galaxy redshift survey \cite{fedli2009}, cosmic 
microwave background radiation (CMBR) data \cite{caldwell2004, huang2006} convincingly suggest that 
the rate of expansion of our universe is positive. i.e. We live in an accelerating expanding universe. In 2006, 
Caldwell et al. \cite{caldwell2006a} and later on Yadav \cite{yadav2012} have given the idea of transitioning 
universe from decelerating phase to accelerating one. However the cause of this sudden transition and 
source of accelerated expansion is still unknown. The most surprising and counterintuitive result coming from 
these observations is the fact that only $\sim 4\%$ of the total energy density of universe is in the form of 
baryonic matter, $\sim 24\%$ is non-baryonic matter and almost $\sim 72\%$ is completely unknown component with negative 
pressure. In the literature, the component with negative pressure is named as dark energy (DE) that produces repulsive 
force which gives rise to the current accelerating expansion of universe. In the recent years, the probable candidate 
of DE such as cosmological constant $(\Lambda)$ \cite{pad2003}, quintessence \cite{martin2008}, phantom \cite{alam2004} 
and chaplygin \cite{bento2002} etc have been proposed. $\Lambda$ is the simplest and most efficient way to generate 
observed accelerated background expansion of universe but it suffers fine tunning and cosmic coincidence problems. This 
make sense to consider some alternative dark energy models.\\   

Now, in the most of the cosmological research, the equation of state of DE $(\omega^{de})$ has been taken 
as time dependent \cite{akarsu2010, kumar2011, kumar2011a}. A Bianchi I model, being the straightforward 
generalization of flat FRW model, is one of the simplest model of anisotropic universe that describes 
the existence of anisotropy in the early stage of evolution of universe. 
In 2008, Rodrigues \cite{rod2008} has developed $\Lambda$CDM 
model in Bianchi I space-time. Later on, Kumar and Singh \cite{kumar2011} have studied the minimal interaction of 
perfect fluid and anisotropic DE by applying special law of variation of Hubble's parameter that 
yields constant value of deceleration parameter (DP). This study leads to 
the two type of cosmological models - singular and non singular. 
However, in the present work, we have searched a model of transitioning universe with singular origin. 
Amendola \cite{amendola2003} and Riess et al \cite{riess2001} have investigated that acceleration of universe 
is accelerating at present epoch, but it was decelerating in the past i.e. there has been transition of universe from 
early decelerating phase to current accelerating phase. Various DE models with time dependent $\omega^{(de)}$ and 
skewness parameters for homogeneous Bianchi cosmologies exist \cite{yadav2011, yadav2011a, yadav2011b}. But here 
we present generalized form of hybrid expansion law (HEL) for scale factor in Bianchi - I space-time with anisotropic DE 
and perfect fluid by using an approach similar to Kumar and Singh \cite{kumar2011}. However, it is to be noted that 
explicit expressions of cosmological parameters are new and different from the other authors. \\

In this paper, we study model of transitioning universe which explain accelerating universe as well as 
singularity free universe, in Bianchi I space-time. To our knowledge, 
this is the first study of anisotropic DE with variable DP. The outline of the 
paper is as follows: In section 2, metric and field equations are described. HEl cosmology and solution of 
field equations along with physical features have been sought for in section 3. Finally, the conclusions are 
summarized in section 4.\\  
\section{Field equations}
The spatially homogeneous and anisotropic Bianchi-I space-time is described by the line element
\begin{equation}
\label{ref1}
 ds^{2}=-dt^{2}+X^{2}(t)dx^{2}+Y^{2}(t)dy^{2}+Z^{2}(t)dz^{2},
\end{equation}
Here, X(t), Y(t), Z(t) are scale factor along x, y and z-direction.\\

Einstein's field equations in case of a mixture of perfect fluid and anisotropic DE components in the gravitational units 
read as
\begin{equation}
 R_{ij}-\frac{1}{2}g_{ij}R = -T_{ij}^{(m)} - T_{ij}^{(de)} ,
\end{equation}
where $T_{ij}^{(m)}$ and $T_{ij}^{(de)}$ are the energy momentum tensors of perfect fluid and 
DE respectively. These are given by
\begin{equation}
 \label{emtp}
T_{ij}^{(m)} = diag[-\rho^{(m)}, p^{(m)}, p^{(m)}, p^{(m)}],
\end{equation}
and
\[
T_{ij}^{(de)} = diag [-\rho^{(de)}, p^{(de)}, p^{(de)}, p^{(de)}].
\]
\begin{equation}
 \label{emtd}
= diag [-1, \omega + \delta, \omega + \gamma, \omega + \eta]\rho^{(de)},
\end{equation}
where $p^{(m)}$, $\rho^{(m)}$, $\rho^{(de)}$ are pressure, energy density of perfect fluid and DE components respectively. 
$\omega$ is the EoS parameter of DE; $\delta(t)$, $\gamma(t)$ and $\eta(t)$ are skewness parameters, which modify EoS of 
DE component along the spatial directions. \\ 
we define $a= (XYZ)^{\frac{1}{3}}$ as average scale factor and H is Hubble's 
parameter such that
\begin{equation}
 H=\frac{\dot{a}}{a},
\end{equation}
where an over dot denotes derivative with respect to the cosmic time $t$.
For the line element (1), the field equation (2) can be written as
\begin{equation}
\label{fe1}
 \frac{\ddot{Y}}{Y}+\frac{\ddot{Z}}{Z}+\frac{\dot{Y}\dot{Z}}{YZ} = -p^{(m)}-(\omega + \delta)\rho^{(de)},
\end{equation}
\begin{equation}
\label{fe2}
 \frac{\ddot{Z}}{Z}+\frac{\ddot{X}}{X}+\frac{\dot{X}\dot{Z}}{XZ} = -p^{(m)}-(\omega + \gamma)\rho^{(de)},
\end{equation}
\begin{equation}
\label{fe3}
 \frac{\ddot{X}}{X}+\frac{\ddot{Y}}{Y}+\frac{\dot{X}\dot{Y}}{XY} = -p^{(m)}-(\omega + \eta)\rho^{(de)},
\end{equation}
\begin{equation}
\label{fe4}
\frac{\dot{X}\dot{Y}}{XY}+\frac{\dot{Y}\dot{Z}}{YZ}+\frac{\dot{Z}\dot{X}}{ZX} = \rho^{(m)} + \rho^{(de)}.
\end{equation}
The above equation can also be written as
\begin{equation}
\label{fe5}
p^{(m)} + \frac{1}{3}(3\omega + \delta + \gamma + \eta)\rho^{(de)} = H^{2}(2q-1)-\sigma^{2},
\end{equation}
\begin{equation}
 \label{fe6}
\rho^{(m)} + \rho^{(de)} = 3H^{2} - \sigma^{2},
\end{equation}
Here, $q$, $\sigma$ are deceleration parameter and shear scalar respectively.\\
Following, Kumar and Singh \cite{kumar2011}, Yadav \cite{yadav2011b}, we assume that 
perfect fluid and DE components interact minimally. Therefore the energy momentum tensors of the 
two sources may be conserved separately.\\
The energy conservation equations $T^{(de)ij}_{;j} = 0$, of DE components leads to 
\begin{equation}
\label{ec1}
\dot{\rho}^{(de)}+3\rho^{(de)}(\omega+1)H + \rho^{(de)}(\delta H_{x} + \gamma H_{y} + \eta H_{z}) = 0.
\end{equation}
whereas the energy conservation equations $T^{(m)ij}_{;j} = 0$, of perfect fluid yields 
\begin{equation}
\label{ec2}
\dot{\rho}^{(m)} + 3(\rho^{(m)} + p^{(m)})H = 0,
\end{equation}
Here, we have used the equation of state $p^{(de)} = \omega \rho^{(de)}$.\\
\section{HEL cosmology}
Following, Yadav \cite{yadav2012}, Yadav and Sharma \cite{yadav2013} and Yadav et al \cite{yadav2015}, we 
consider the generalized HEL for scale factor as following
\begin{equation}
 \label{scale}
a = (t^{n}e^{kt})^{\frac{1}{m}},
\end{equation}
where $m$, $n$ and $k$ are non negative constant. The proposed law gives the time varying 
deceleration parameter which describes the transitioning universe.\\

In this paper, our aim is to describes the anisotropic nature of dynamical dark energy by HEL cosmology. So, 
in order to study the anisotropic DE, we have to assume that skewness parameters are time dependent quantities. 
At the same time we have to ensure that the time dependent form of skewness parameters provide exact solution of field 
equations (\ref{fe1})-(\ref{fe4}) together with (\ref{scale}). We find that the following time dependent 
forms of skewness parameters are appropriate to meet the requisite task:
\begin{equation}
 \label{skew1}
\delta(t) = \alpha(H_{y} + H_{z}),
\end{equation}
\begin{equation}
 \label{skew2}
\gamma(t) = \eta(t) = -\alpha H_{x},
\end{equation}
where $\alpha$ is an arbitrary constant, which parameterizes the anisotropy of DE. In the literature, many authors 
have considered anisotropy DE with constant DP \cite{kumar2011, yadav2011b}.\\

In view of assumptions (\ref{skew1}), (\ref{skew2}), one can solve equation (\ref{ec1}) as
\begin{equation}
 \label{rho^{de}}
\rho^{(de)} = \rho^{(de)}_{0}a^{-3(\omega + 1)},
\end{equation}
where $\rho_{0}^{(de)}$ is a positive constant of integration.\\
Solving equations (\ref{fe1}) - (\ref{fe4}) together with (\ref{scale}), we get
\begin{equation}
 X(t) = a_{1}(t^{n}e^{kt})^{1/m}exp\left[b_{1}\int(t^{n}e^{kt})^{-3/m}dt - \frac{2\alpha \rho_{0}^{(de)}}{3\omega}
\int(t^{n}e^{kt})^{-3(\omega+1)/m}dt\right],
\end{equation}
\begin{equation}
 Y(t) = a_{2}(t^{n}e^{kt})^{1/m}exp\left[b_{2}\int(t^{n}e^{kt})^{-3/m}dt + \frac{\alpha \rho_{0}^{(de)}}{3\omega}
\int(t^{n}e^{kt})^{-3(\omega+1)/m}dt\right],
\end{equation}
\begin{equation}
 Z(t) = a_{3}(t^{n}e^{kt})^{1/m}exp\left[b_{3}\int(t^{n}e^{kt})^{-3/m}dt + \frac{\alpha \rho_{0}^{(de)}}{3\omega}
\int(t^{n}e^{kt})^{-3(\omega+1)/m}dt\right],
\end{equation}
Here $a_{1}$, $a_{2}$, $a_{3}$ $\&$ $b_{1}$, $b_{2}$, $b_{3}$ are constant satisfying the following condition.
$$a_{1}a_{2}a_{2} = 1$$ 
$$b_{1}+b_{2}+b_{3} = 0$$.
The DP, in the derived model is given by
\begin{equation}
 \label{dp}
q = \frac{d}{dt}\left(\frac{1}{H}\right) - 1 = \frac{mn}{(n+kt)^2} -1.
\end{equation}

Equation (\ref{dp})  exhibits the time dependency of DP. A negative value of $q$ corresponds to 
accelerating universe while positive $q$ indicates the decelerating phase of universe. For $m \textgreater n$ and $t = 0$,
the universe is in decelerating phase while with passage of time it turns into accelerating phase. The transition of 
universe from decelerating to accelerating phase take place at $t = \frac{\sqrt{mn}-n}{k}$. For $t \rightarrow \infty$, 
we get $q = -1$; incidentally this value of DP leads to $\frac{dH}{dt} = 0$, which implies the greatest value of 
Hubble's parameter and the fastest rate of expansion of the universe.\\
 
The physical parameters such as directional Hubble parameters $H_{x}$, $H_{y}$, $H_{z})$, average Hubble parameter $(H)$, 
expansion scalar $(\theta)$, spatial volume $(V)$ and mean anisotropy parameter $(A_{m})$ are, respectively given by
\begin{equation}
\label{hx}
H_{x} = \frac{1}{m}\left(\frac{n}{t}+k\right)+b_{1}(t^{n}e^{kt})^{-\frac{3}{m}}-\frac{2\alpha \rho^{(de)}_{0}}{3\omega}
(t^{n}e^{kt})^{-\frac{3(\omega+1)}{m}}.
\end{equation}
\begin{equation}
\label{hy}
H_{y} = \frac{1}{m}\left(\frac{n}{t}+k\right)+b_{2}(t^{n}e^{kt})^{-\frac{3}{m}}+\frac{\alpha \rho^{(de)}_{0}}{3\omega}
(t^{n}e^{kt})^{-\frac{3(\omega+1)}{m}}.
\end{equation}
\begin{equation}
\label{hz}
H_{z} = \frac{1}{m}\left(\frac{n}{t}+k\right)+b_{3}(t^{n}e^{kt})^{-\frac{3}{m}}+\frac{\alpha \rho^{(de)}_{0}}{3\omega}
(t^{n}e^{kt})^{-\frac{3(\omega+1)}{m}}.
\end{equation} 
\begin{equation}
 \label{h}
H = \frac{1}{m}\left(\frac{n}{t}+k\right).
\end{equation}
\begin{equation}
\label{theta}
\theta = \frac{3}{m}\left(\frac{n}{t}+k\right).
\end{equation}
\begin{equation}
\label{volume}
V = (t^{n}e^{kt})^{\frac{3}{m}}.
\end{equation}
\begin{equation}
 \label{anisotropy}
A_{m} = \frac{2m^{2}}{3\left(\frac{n}{t}+k\right)}\left[\beta(t^{n}e^{kt})^{-\frac{6}{m}} + \frac{\alpha^{2}
\rho_{0}^{(de)^{2}}}{3\omega^{2}}(t^{n}e^{kt})^{-\frac{6(\omega+1)}{m}} - \frac{b_{1}\alpha\rho_{0}^{(de)}}
{\omega}(t^{n}e^{kt})^{-\frac{3(\omega+2)}{m}}\right],
\end{equation}
where $\beta = b_{1}^{2} + b_{2}^{2} + b_{1}b_{2}$.\\
\begin{figure*}[thbp]
\begin{tabular}{rl}
\includegraphics[width=9cm]{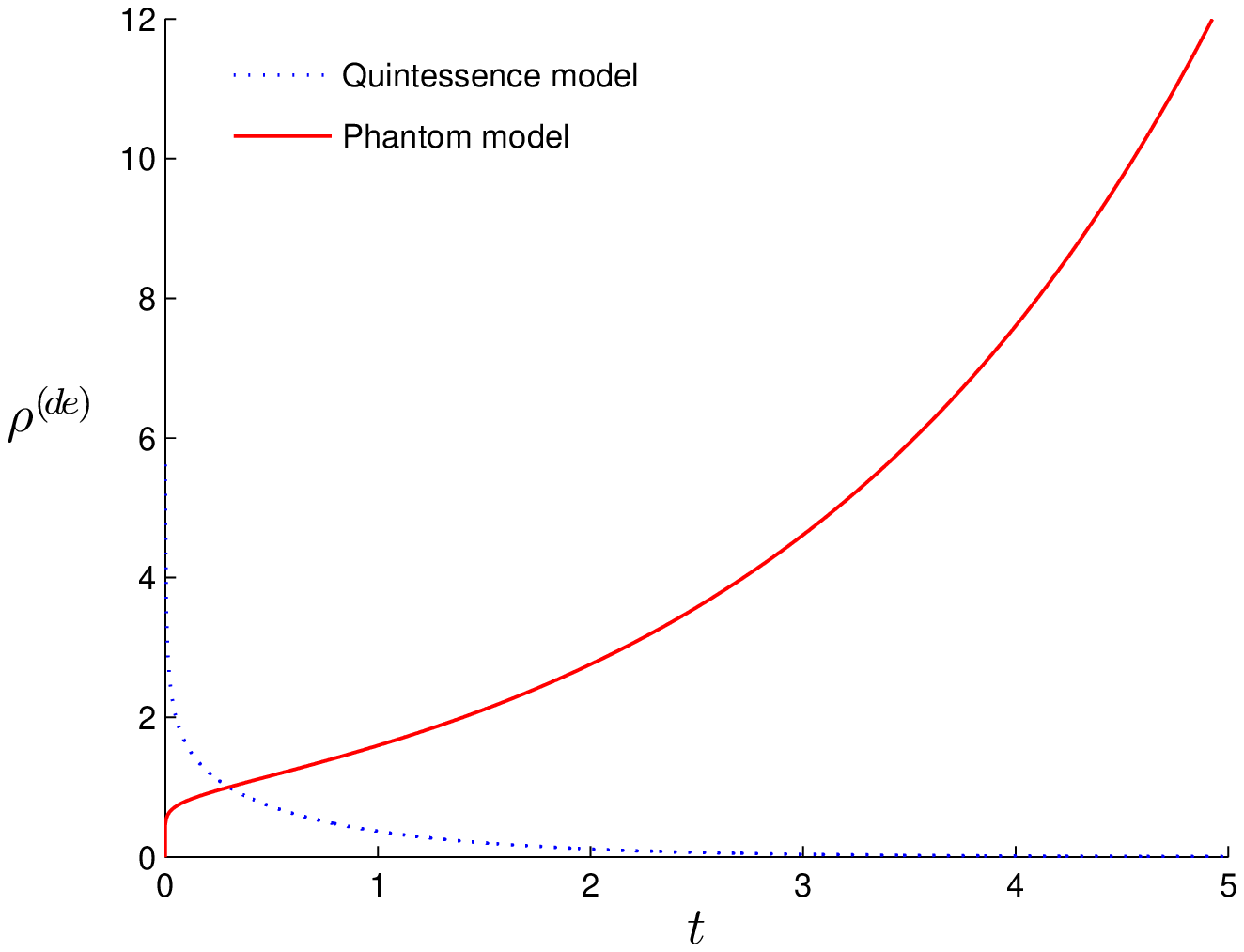}
\includegraphics[width=9cm]{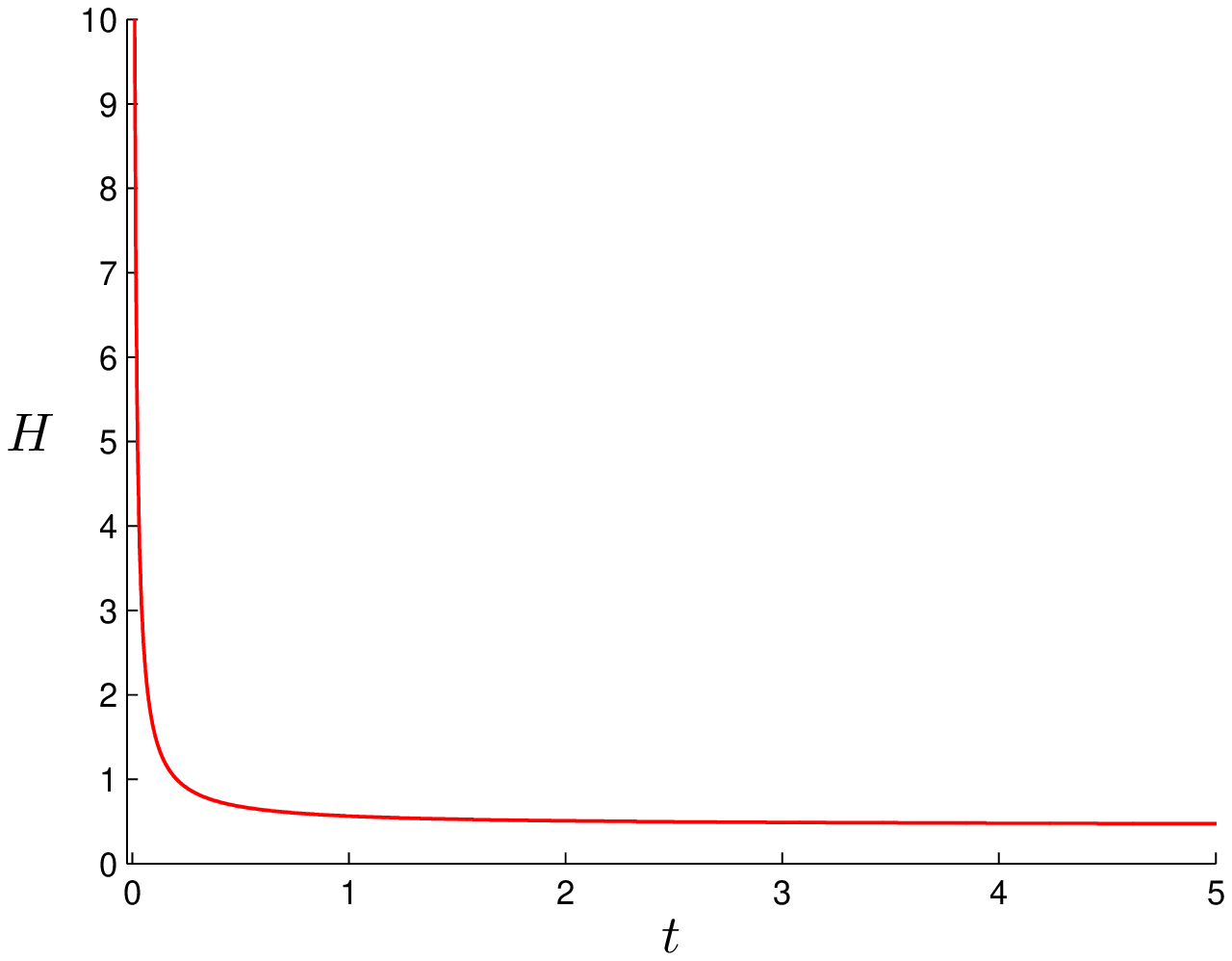}\\
\includegraphics[width=9cm]{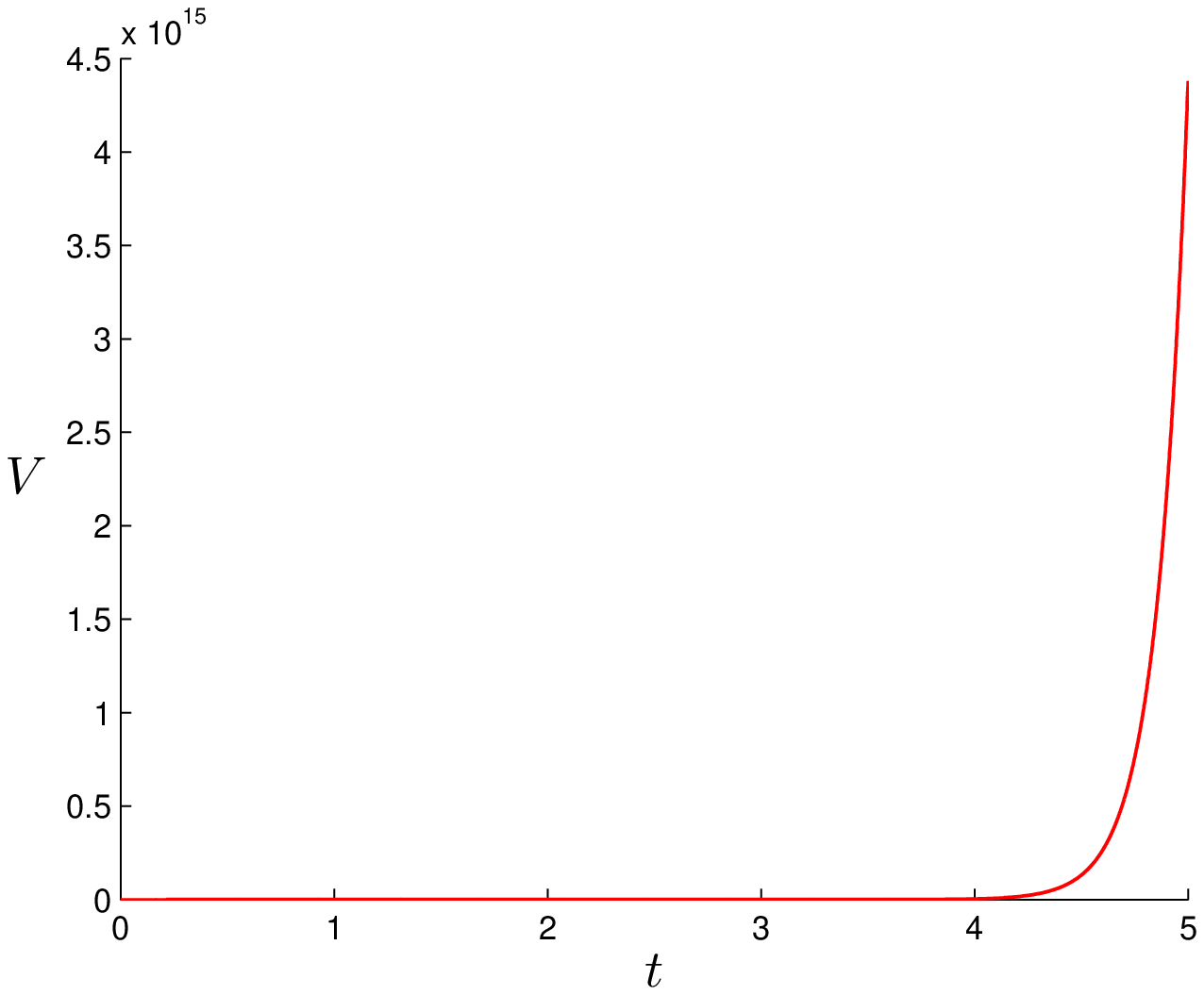}
\includegraphics[width=9cm]{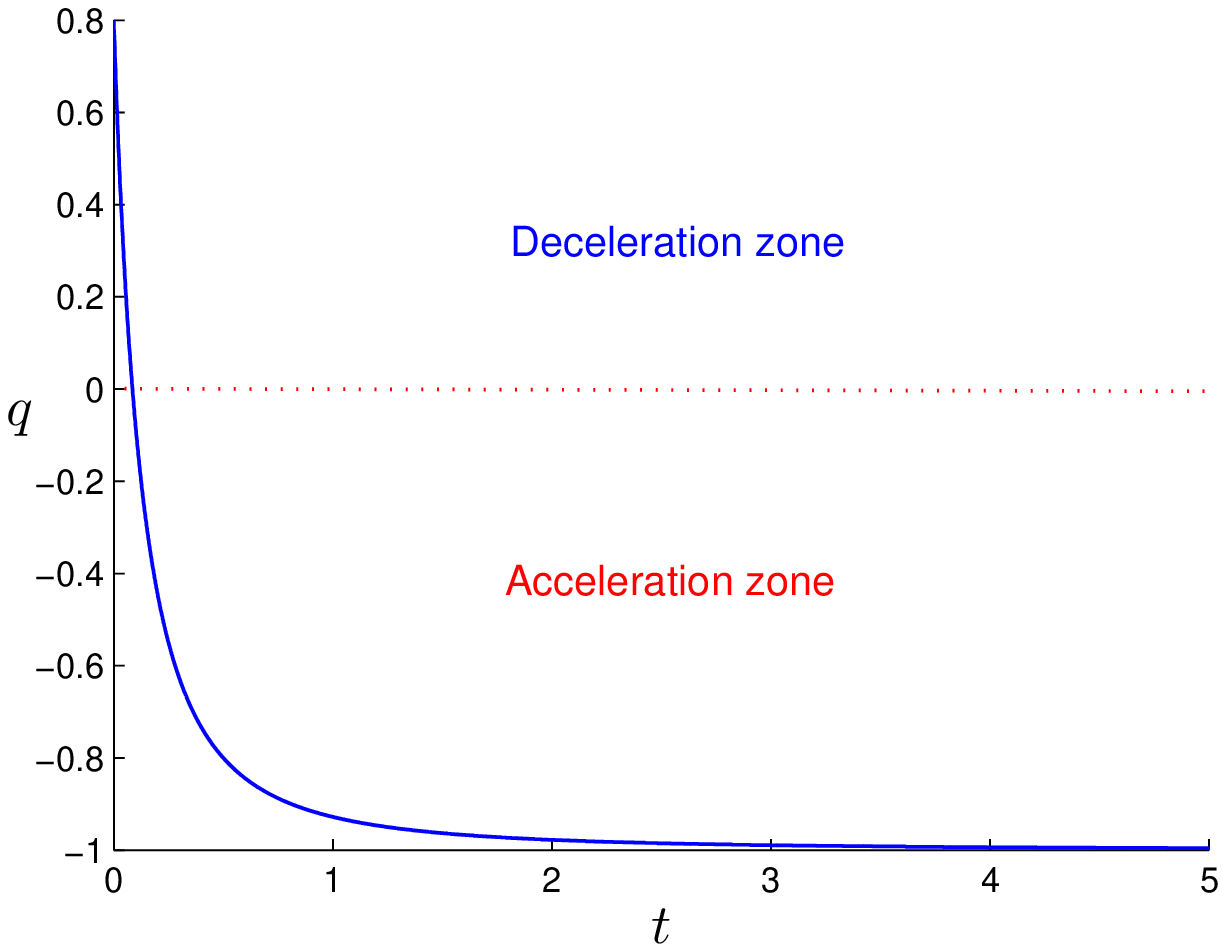}
\end{tabular}
\caption{Variation of dark energy density (upper left panel), mean Hubble's parameter (upper right panel),
volume (lower left panel) and DP (lower right panel) versus time.}
\end{figure*}
The shear scalar of the model is given by
\begin{equation}
 \label{shear}
\sigma^{2} = \beta(t^{n}e^{kt})^{-\frac{6}{m}} + \frac{\alpha^{2}
\rho_{0}^{(de)^{2}}}{3\omega^{2}}(t^{n}e^{kt})^{-\frac{6(\omega+1)}{m}} - \frac{b_{1}\alpha\rho_{0}^{(de)}}
{\omega}(t^{n}e^{kt})^{-\frac{3(\omega+2)}{m}}.
\end{equation}
The energy density and pressure of DE components are given by
\begin{equation}
 \label{rho(de)}
\rho^{(de)} = \rho_{0}^{(de)}(t^{n}e^{kt})^{-\frac{3(\omega + 1)}{m}}.
\end{equation}
\begin{equation}
 \label{p(de)}
p^{(de)} = \omega \rho_{0}^{(de)}(t^{n}e^{kt})^{-\frac{3(\omega + 1)}{m}}. 
\end{equation}
The skewness parameters of DE are 
obtained as 
\begin{equation}
 \label{delta}
\delta(t) = \alpha\left[\frac{2}{m}\left(\frac{n}{t}+k\right)-b_{1}(t^{n}e^{kt})^{-\frac{3}{m}} + 
\frac{2\alpha \rho_{0}^{(de)}}{3\omega}(t^{n}e^{kt})^{-\frac{3(\omega+1)}{m}}\right].
\end{equation}
\begin{equation}
 \label{gamma}
\gamma(t) = \eta(t) = -\alpha\left[\frac{1}{m}\left(\frac{n}{t}+k\right) + b_{1}(t^{n}e^{kt})^{-\frac{3}{m}} - 
\frac{2\alpha \rho_{0}^{(de)}}{3\omega}(t^{n}e^{kt})^{-\frac{3(\omega+1)}{m}}\right].
\end{equation}
In the light of above solution, equations (\ref{fe5}) and (\ref{fe6}) lead to
\[
 p^{(m)} = \frac{1}{m^{2}}\left(\frac{2mn}{(n+kt)^{2}}-3\right)\left(\frac{n}{t}+k\right)^{2} - \beta(t^{n}e^{kt})^
{-\frac{6}{m}} - \frac{\alpha^{2}\rho_{0}^{(de)}}{3\omega^{2}}(t^{n}e^{kt})^{-\frac{6(\omega+1)}{m}} +
\]
\begin{equation}
 \label{p(m)} 
\frac{b_{1}\alpha\rho_{0}^{(de)}}{\omega}(t^{n}e^{kt})^{-\frac{3(\omega+2)}{m}}-\omega -\frac{b_{1}\alpha}{3}(t^{n}e^{kt})
^{-\frac{3}{m}} + \frac{2\alpha\rho_{0}^{(de)}}{9\omega}(t^{n}e^{kt})^{-\frac{3(\omega+1)}{m}}.
\end{equation}
\begin{equation}
\label{rho(m)}
\rho^{(m)} = \frac{3}{m^{2}}\left(\frac{n}{t}+k\right)^{2}-\beta(t^{n}e^{kt})^{-\frac{6}{m}} - \frac{\alpha^{2}\rho_{0}^{2}}
{3\omega^{2}}(t^{n}e^{kt})^{-\frac{6(\omega +1)}{m}} + \omega\rho_{0}^{(de)}(t^{n}e^{kt})^{-\frac{3(\omega +2)}{m}}
\left(\frac{b_{1}\alpha}{\omega^{2}} - (t^{n}e^{kt})^{\frac{1}{m}}\right).
\end{equation}
\begin{figure*}[thbp]
\begin{tabular}{rl}
\includegraphics[width=15cm]{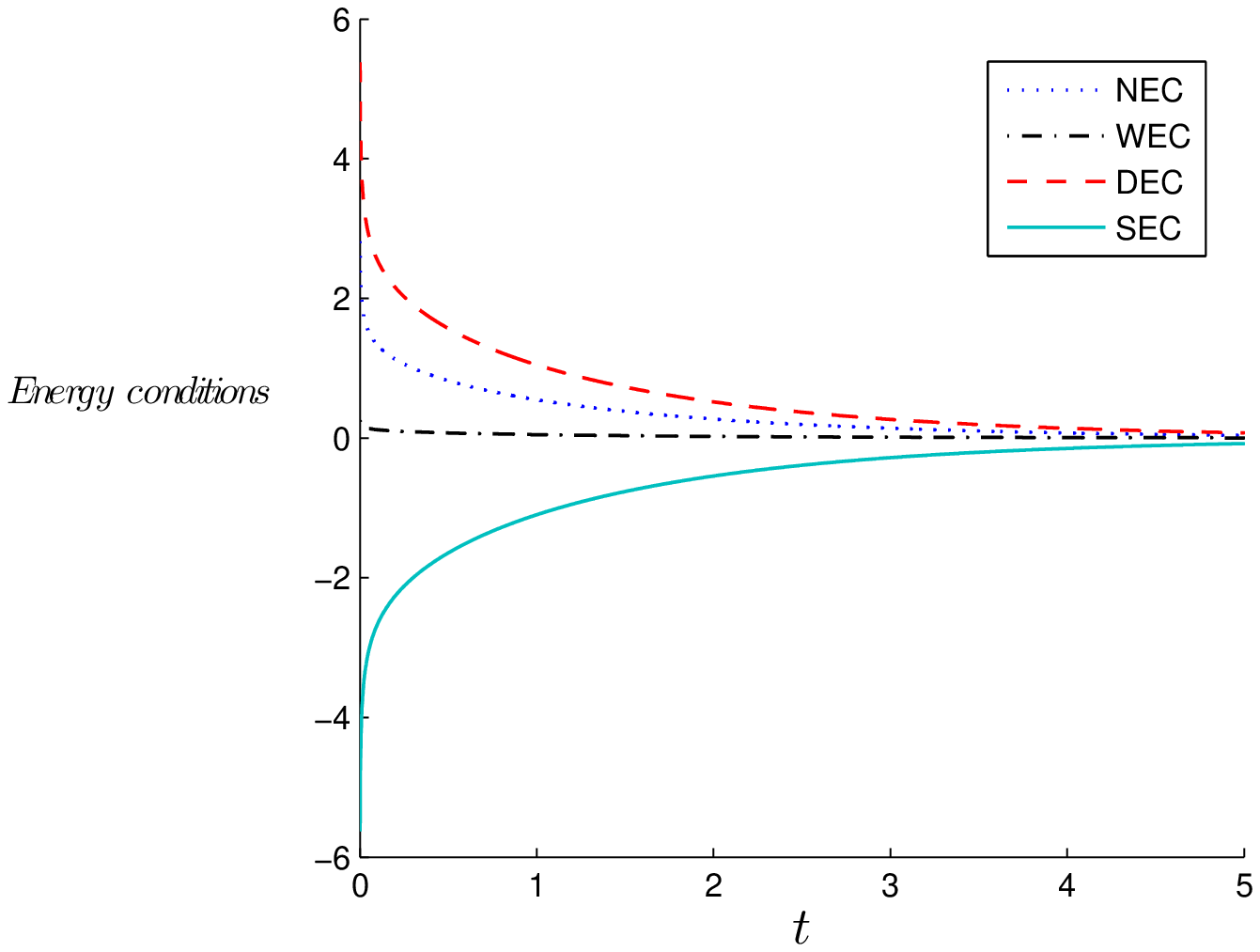}\\
\includegraphics[width=15cm]{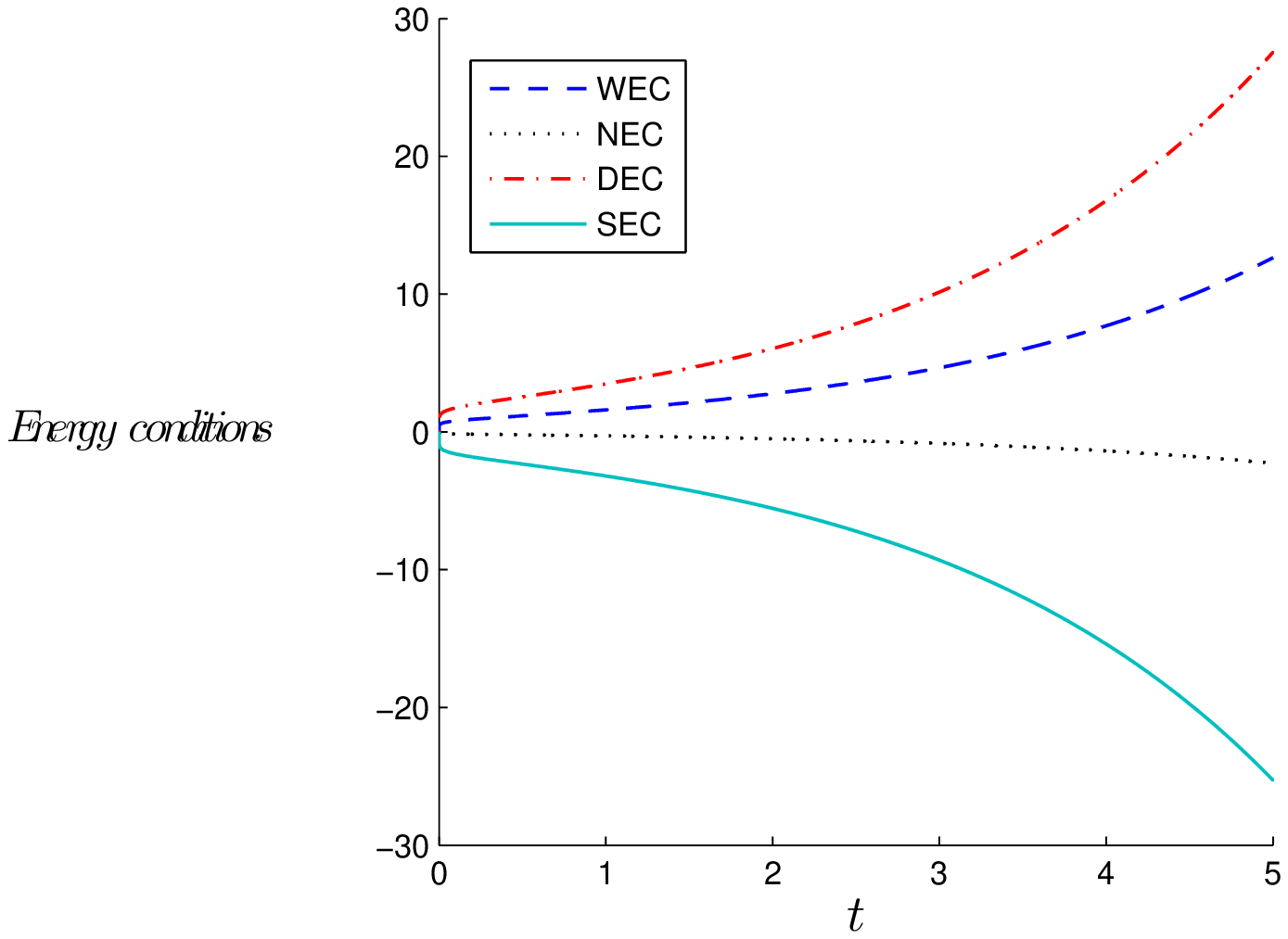}
\end{tabular}
\caption{Single plot of energy conditions for Quintessence model (upper panel) and Phantom model (lower panel).}
\end{figure*}

The spatial scale factors $(X(t), Y(t), Z(t)$ vanish at $t = 0$. Therefore, the model has point type 
singularity at $t = 0$. However $n = 0$ generates the singularity free model of universe. The difference between 
axial skewness parameters and hence EoS parameter along x-axis and y-axis (or z-axis) is $\frac{3\alpha}{m}\left(
\frac{n}{t}+k\right)$, which decreasing function of time. Hence, the anisotropy of DE decreases as time increases and 
finally disappear at $t \rightarrow \infty$. One can constrain the geometry of universe by choosing appropriate value of 
problem parameters such as for $a_{2} = a_{3}$ and $b_{2} = b_{3}$, the universe has prolate geometry as 
$X(t) > Y(t) = Z(t)$.\\

In figure panel 1, we graphed the cosmological parameters of derived model against $t$. The behaviour is quite evident: 
the dark energy density of quintessence model decreases while for phantom model, it increases with passage of time; the 
spatial volume $(V)$ is increasing function of time which shows the expansion of universe; the behaviour of $q$ against 
$t$ describes the features of transitioning universe.\\

The left hand side of energy conditions for quintessence and phantom model have been graphed in figure panel 2. 
It is interesting to note here that null energy condition (NEC) as well as dominant energy condition (DEC) are satisfied 
for both the models - quintessence and phantom model. These models do not satisfy the strong energy condition (SEC), as 
expected. The time dependency of skewness parameters give rise to an ellipsoidality of the universe in spite of 
the inflation.  
\section{Conclusion}
In this paper, we have searched a model of accelerating universe as well singularity free universe with minimal 
interaction between perfect fluid and anisotropic DE in Bianchi I space-time. The anisotropic DE has dynamical 
energy density. It is to be noted that our procedure of solving differential equations are similar to Kumar and Singh 
\cite{kumar2011} but the explicit expressions of cosmological parameters are all together different. In fact, the solution 
presented in this paper generalize the solution obtained by Kumar and Singh \cite{kumar2011}. It is observed that 
$k = 0$, corresponds to power law expansion that seems to describe dynamics of universe from big bang to present epoch 
while $n = 0$ seems reasonable to
 project singularity free universe. In general, the model has point type singularity at 
initial epoch as scale factors and volume vanish at $t = 0$. The pressure of DE as well as perfect 
fluid become negligible whereas the scale factors and volume becomes sufficiently large as $t \rightarrow \infty$. 
All figures depict interesting features of the present cosmological model in terms of DE and 
other physical parameters. The derived models violate the SEC while WEC, NEC and DEC may be preserved in quintessence 
model which can be acceptable in present time. Thus the present analysis reveals that 
quintessence model is suitable for describing the accelerating nature of universe consistent with the observations.  
One important feature of the model is that it gives the possibility of accelerating universe with positive pressure 
in absence of DE which need to be tested by other model/theory. However, it is to be noted that 
our model is based on alternative candidate of DE other than cosmological terms $(\Lambda)$ and does not invoke 
any other agent or theory for absence of DE.


\begin{thebibliography}{99}
\bibitem{riess1998} Riess, A. G., et al.: Astrophys. J. \textbf{116}, 1009 (1998)
\bibitem{perlmutter1999} Perlmutter, S., et al.: Astrophys. J. \textbf{517}, 565 (1999)
\bibitem{fedli2009} Fedli, C., Moscardini, L., Bertelmann,  M.: Astron. Astrophys. \textbf{500}, 667 (2009)
\bibitem{caldwell2004}  Caldwell, R. R., Doran, M.: \textbf{69}, 103517 (2004)
\bibitem{huang2006} Huang, Z-Yi. et al.: \textbf{JCAP} 05, 013 (2006)
\bibitem{caldwell2006a} Caldwell, et al.: Phys. Rev. D \textbf{73}, 023513 (2006)
\bibitem{yadav2012} Yadav, A. K.: Res. Astron. Astrophys. \textbf{12}, 1467 (2012)   
\bibitem{pad2003} Padmanabhan, T.: Phys. Rept. \textbf{380}, 225 (2003)
\bibitem{martin2008} Martin, J.: Mod. Phys. Lett. A \textbf{23}, 1252 (2008)
\bibitem{alam2004} Alam, U. et al.: Mon. Not. R. Astron Soc. \textbf{354}, 275 (2004)
\bibitem{bento2002} Bento, M.C., Bertolami, O., Sen, A. A.: Phys. Rev. D \textbf{66}, 043507 (2002)
\bibitem{akarsu2010} Akarsu, \"{O}., Killinc, C. B.: Gen. Relativ. Grav. \textbf{42}, 119 (2010)
\bibitem{kumar2011} Kumar, S., Singh, G. P.: Gen. Relativ. Grav. \textbf{43}, 1427 (2011)
\bibitem{kumar2011a} Kumar, S., Yadav, A. K.: Mod. Phys. Lett. A \textbf{26}, 647 (2011)
\bibitem{rod2008} Rodrigues, D. C.: Phys. Rev. D \textbf{77}, 023534 (2008)
\bibitem{amendola2003} Amendola, L.: Mon. Not. Royal Astron. Soc. \textbf{342}, 221 (2003)
\bibitem{riess2001} Riess, A. G., Nugent, P. E., Gilliland, R. L. et al.: APJ \textbf{560}, 49 (2001)
\bibitem{yadav2011} Yadav, A. K., Yadav, L.: Int. J. Theor. Phys. \textbf{50}, 218 (2011)
\bibitem{yadav2011a} Yadav, A. K., Rahaman, F., Ray, S.: Int. J. Theor. Phys. \textbf{50}, 871 (2011)
\bibitem{yadav2011b} Yadav, A. K.: Astrophys. Space. Sc. \textbf{335}, 565 (2011)
\bibitem{yadav2013} Yadav, A. K., Sharma, A.: Res. Astron. Astrophys. \textbf{13}, 501 (2013)
\bibitem{yadav2015} Yadav, A. K., Srivastava, P. K., Yadav, L.: Int. J. Theor. Phys. \textbf{54}, 1671 (2015)  
\end{thebibliography}
\end{document}